\documentclass[prl,twocolumn,superscriptaddress]{revtex4-1}
\usepackage{amsmath,amssymb,mathrsfs}
\usepackage{graphicx,import}
\usepackage[usenames,dvipsnames]{color}
\usepackage[colorlinks=true,linkcolor=blue,citecolor=BrickRed,urlcolor=blue]{hyperref}

\def\be{\begin{equation}}      
\def\ee{\end{equation}}
\def\bea{\begin{eqnarray}}      
\def\eea{\end{eqnarray}}

\def\tr{\mbox{tr}}
\def\Br{\mathbf{r}}
\def\Bs{\mathbf{S}}
\def\Bq{\mathbf{Q}}

\begin{document}

\title{Frustrated Magnetism of Dipolar Molecules on a Square Optical Lattice: Prediction of a Quantum Paramagnetic Ground State}

\author{Haiyuan Zou}
\affiliation{Wilczek Quantum Center, School of Physics and Astronomy and T. D. Lee Institute,  Shanghai Jiao Tong University, Shanghai 200240, China}
\affiliation{Department of Physics and Astronomy, University of Pittsburgh, Pittsburgh, Pennsylvania 15260, USA}

\author{Erhai Zhao}
\affiliation{Department of Physics and Astronomy, George Mason University, Fairfax, Virginia 22030, USA}

\author{W. Vincent Liu}
\affiliation{Wilczek Quantum Center, School of Physics and Astronomy and T. D. Lee Institute,  Shanghai Jiao Tong University, Shanghai 200240, China}
\affiliation{Department of Physics and Astronomy, University of Pittsburgh, Pittsburgh, PA 15260}
\affiliation{Center for Cold Atom Physics, Chinese Academy of Sciences, Wuhan 430071, China}

\begin{abstract}
Motivated by the experimental realization of quantum spin models of polar molecule KRb in optical lattices,
we analyze the spin 1/2 dipolar Heisenberg model with competing anisotropic, long-range exchange interactions.
We show that, by tilting the orientation of dipoles using an external electric field, the dipolar spin system on square lattice 
comes close to a maximally frustrated region similar, but not identical, to that of the $J_1$-$J_2$ model.
This provides a simple yet powerful route to potentially realize a quantum spin liquid without the need for a triangular or kagome lattice. 
The ground state phase diagrams obtained from Schwinger-boson and spin-wave theories consistently show
a spin disordered region between the N$\acute{\textrm{e}}$el, stripe, and spiral phase. 
The existence of a finite quantum paramagnetic region is further confirmed by an unbiased variational ansatz
based on tensor network states and a tensor renormalization group.
\end{abstract}
\maketitle

Understanding highly entangled quantum matter remains a challenging goal of condensed matter physics~\cite{newRev2}.
One paradigmatic example is quantum spin liquids in frustrated spin systems which defy any conventional long range order characterized by broken symmetry at zero temperature~\cite{newRev2,newRev,diep2013frustrated}. Instead, the ground state features long-range entanglement and nonlocal excitations.
Spin liquids are also fertile ground for studying quantum phases described by gauge field theories and topological order~\cite{PhysRevB.65.165113}.
While the existence of spin liquids has been firmly established in a number of exactly solvable models, e.g., the toric code~\cite{Kitaev20032}
or the honeycomb Kitaev model~\cite{Kitaev20062}, the nature of the ground states for many frustrated spin models, e.g., the Heisenberg model on kagome 
lattices or the $J_1$-$J_2$ model on square lattices, still remains controversial despite the great theoretical progress in recent years~\cite{Yan1173,Gapless2013,J1J2Jiang,J1J2Plaq,tnj1j2}.  
An unambiguous experimental identification of quantum spin liquids in solid state materials also seems elusive~\cite{newRev2}. 
It is, then, important to explore new physical systems that can cleanly realize
well-defined spin models which have potential spin liquid ground states.

Recent breakthrough experiments on magnetic atoms \cite{PhysRevLett.111.185305} and polar molecules ~\cite{Yan:2013xe,PhysRevLett.113.195302} confined in deep optical lattices introduced 
a new class of lattice spin models with competing exchange interactions that are long-ranged and anisotropic. 
The resulting spin Hamiltonians, such as the dipolar $XXZ$
and dipolar Heisenberg models, are highly tunable by the external fields that couple to the magnetic and electric dipoles~\cite{PhysRevLett.107.115301,DSL2015}. 
Here, we show that these models on square lattices
feature strong exchange (not geometric) frustration and a quantum paramagnetic ground state for intermediate dipole tilting angles. This claim is consistently supported by physical arguments, two independent semiclassical analytical methods, and full numerical calculation based on tensor network ansatz~\cite{PEPS1,iPEPS,iPEPS2,LevinTRG,HOTRG1}.
Our key insight is that spin liquids may arise naturally from the system of tilted, interacting dipoles on square lattices, without the
requirement of peculiar (e.g., triangular or kagome) lattices or exotic (e.g., Kitaev or ring-exchange) interactions. 

{\it The dipolar XXZ and Heisenberg model.}---First, we define the dipolar $XXZ$ model on a square optical lattice,
\be 
H_{XXZ}=\frac{J}{2}\sum_{i\neq j} f(\mathbf{r}_i-\mathbf{r}_j) (S^x_i S^x_j+S^y_i S^y_j+\eta S^z_i S^z_j).
\ee
Here $i$ and $j$ label the lattice sites, $\mathbf{S}_i=(S^x_i,S^y_i,S^z_i)$ are the spin (or pseudospin) operators at site $i$, and $\eta$ is the exchange anisotropy.
The key new feature here is that the coupling between the two spins depends on their relative position $\mathbf{r}=\mathbf{r}_i-\mathbf{r}_j$
and the external field (dipole) direction $\hat{d}$

\be
f(\mathbf{r})=[1-3(\hat{r}\cdot \hat{d})^2](a/r)^3,
\ee
with $a$ the lattice constant [Fig. 1(a)]. This geometric factor, characteristic of the dipole-dipole interaction, dictates that spin interactions are long-ranged and anisotropic. For the special case of $\eta=1$, $H_{XXZ}$ reduces to the dipolar Heisenberg model
\be
H_d=\frac{J}{2}\sum_{i\neq j} f(\mathbf{r}_i-\mathbf{r}_j) \mathbf{S}_i \cdot \mathbf{S}_j,
\ee
and for $\eta=0$, it reduces to the dipolar $XY$ model, $H_{XY}$.

Spin models of the form of $H_{XXZ}$ have been realized experimentally in two settings. In Ref. \cite{PhysRevLett.111.185305}, the spin dynamics of a gas of $^{52}$Cr atoms in optical lattices was observed. Each Cr atom carries a magnetic moment of $7\mu_B$ and hyperfine spin $S=3$. An external magnetic field is used to align the magnetic dipoles in the direction of $\hat{d}$. Such a dipolar gas of Cr in a deep lattice is shown to be described by $H_{XXZ}$ with $J=-\mu_0(g\mu_B)^2/4\pi a^3<0$ and $\eta=-2$  \cite{PhysRevLett.111.185305}. 
Note that $J$ induced by the dipolar interaction is, contrary to the superexchange, independent of the tunneling, and it can be set as the unit of energy. 

\begin{figure}
\includegraphics[width=0.48\textwidth]{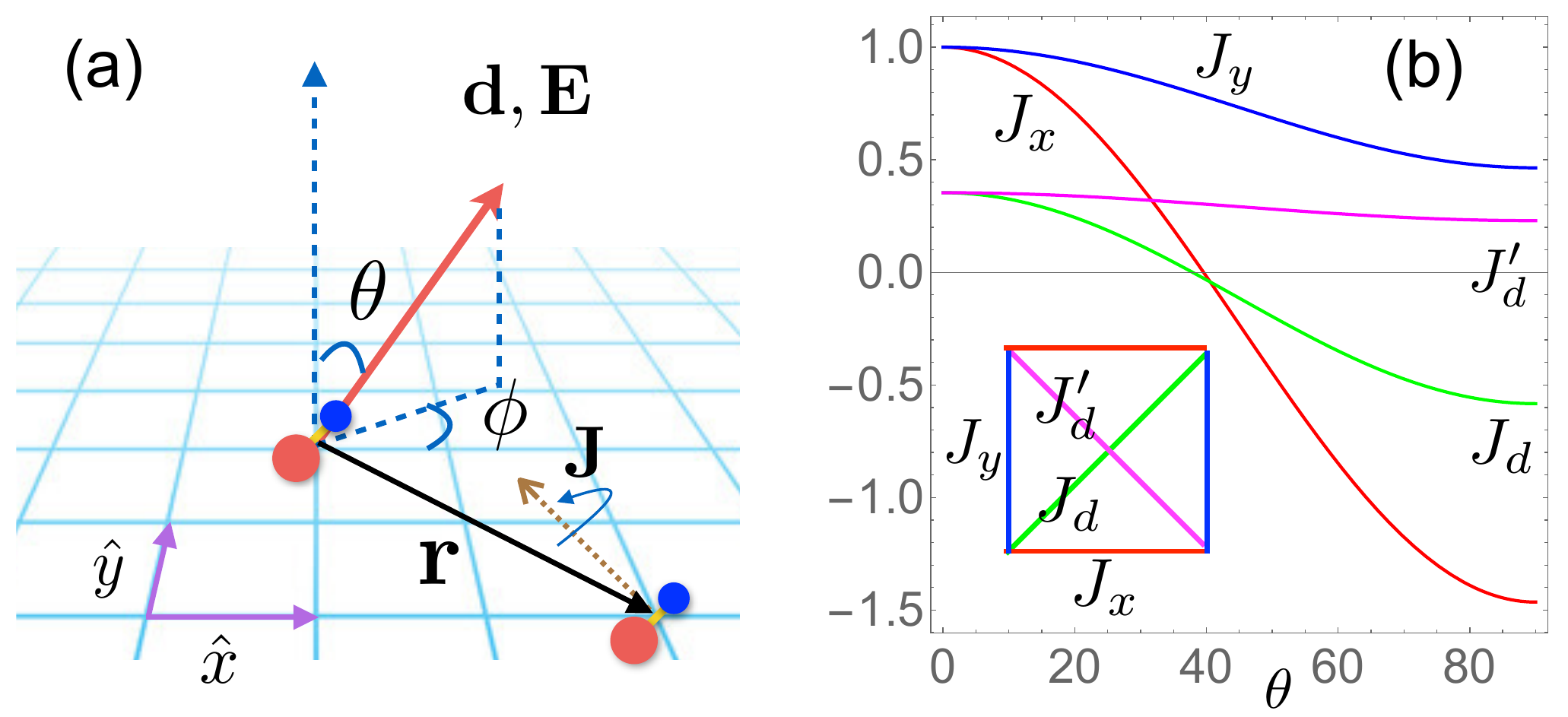}
\caption{(a) Dipolar molecules such as KRb confined in a square optical lattice. The direction of the dipoles $\mathbf{d}$ is tuned by the electric field $\mathbf{E}$. Two rotational states of the molecules play the role of pseudospin up and down. The system  is described by the effective $XXZ$ model Eq. (1). With the proper choice of $E$, it reduces to the dipolar Heisenberg model $H_d$ in Eq. (3). (b) Leading exchange interactions $J_x$, $J_y$, $J_d$, and $J'_d$ (inset) as functions of the dipole tilting angle $\theta$ for fixed $\phi=25^\circ$.
Strong frustration occurs at intermediate $\theta$.} 
\label{modelJ}
\end{figure} 

Polar molecules such as $^{40}$K$^{87}$Rb confined in optical lattices with negligible tunneling provide another way to realize $H_{XXZ}$ with $S=1/2$ and tunable $J$ and $\eta$ \cite{Yan:2013xe}. Each molecule carries an electric dipole moment $\mathbf{d}$ and undergoes rotation with angular momentum $\mathbf{J}$ [see Fig. 1(a)].  
Here, the pseudospin 1/2 refers to two rotational states of the molecule labeled by $|j,m\rangle$, where $j$ is the quantum number of the rotational angular momentum $\mathbf{J}$ and $m$ is its projection onto the quantization axis, chosen as the direction of the external electric field $E$. More details can be found in Ref.~\cite{Yan:2013xe,PhysRevA.84.033619,DSL2015}.
The dipole-dipole interaction projected onto the sub-Hilbert space of the pseudospins then takes the form of a spin Hamiltonian, where the spin flips correspond to transitions between the rotational states.
For example, by choosing $|j,m\rangle=|0,0\rangle$ and $|1,0\rangle$ as the pseudospin down and up respectively, Refs. \cite{PhysRevA.84.033619,DSL2015} showed that the system is described by the effective Hamiltonian $H_{XXZ}$ with
$J=D_t ^2/2\pi\epsilon_0 a^3>0$ and $\eta=(D_1-D_0)^2/2D_t^2>0$. 
Here the dipole matrix element $D_t=\langle1,0|d^0|0,0\rangle$, $D_1=\langle1,0|d^0|1,0\rangle$, $D_0=\langle0,0|d^0|0,0\rangle$, and $d^0$ together with $d^\pm$ form the vector dipole operator in the spherical basis \cite{PhysRevA.84.033619,DSL2015}. 

The anisotropy $\eta$ increases monotonically with $E$. As shown in Ref.~\cite{DSL2015}, 
when $E\simeq 1.7B/|\mathbf{d}|$ with $B$ the energy splitting of the two pseudospin states, $\eta=1$, and one arrives at the dipolar Heisenberg model $H_d$. In the KRb experiment~\cite{Yan:2013xe} carried out at zero field and cubic lattice, $\eta\rightarrow 0$, the dipolar $XY$ model $H_{XY}$ was realized with  $J$ on the order of 100 Hz. Despite the low filling factor and high entropy, coherent spin dynamics was observed via Ramsey spectroscopy \cite{Yan:2013xe} and modeled theoretically in Ref. \cite{PhysRevLett.113.195302}. Recently Yao {\it et al.}~\cite{DSL2015} considered general $\eta$ and worked out the phase diagram of $H_{XXZ}$ on the Kagome and triangular lattice using Density Matrix Renormalization Group (DMRG).. For both lattices, they found evidence for quantum spin liquid centering around the Heisenberg limit, $\eta=1$ and $\theta=0$, in which $\theta$ is defined by $\hat{d}\cdot\hat{x}=\sin\theta\cos\phi$ with $\hat{x}$ representing a base vector of the square lattice. Thus the physics is connected to a geometrically frustrated Heisenberg model on both lattices, with additional longer range interactions and anisotropy $\eta$.   

In this Letter, we study the phases of $H_{d}$ on a square lattice as the dipoles are tilted towards the lattice plane [see Fig. 1(a)] for $S=1/2$ and $J>0$. We show that strong frustration occurs at intermediate dipole tilting angle $\theta$, 
leading to a quantum paramagnetic ground state. We emphasize that, here, the frustration is not imposed by the lattice geometry, but instead, is due to the competition between the exchange interactions, analogous to the $J_1$-$J_2$ model. Relatedly, the quantum paramagnetic phase appears at intermediate $\theta$ values (not around $\theta=0$ as in Ref. \cite{DSL2015}) between the N$\acute{\textrm{e}}$el and the stripe orders. Thus, it differs qualitatively from the spin liquids studied in Ref. \cite{DSL2015}. We will also employ different methods to solve the dipolar quantum spin models. 

\textit{Competing exchanges for tilted dipoles.}---To appreciate the possible phases of $H_{d}$ as $\hat{d}$ is tuned as well as its connection to frustrated quantum spin models \cite{Balents:2010kq,diep2013frustrated}, let us consider the leading exchange couplings between the nearest neighbors, $J_x=Jf(a\hat{x})$ and $J_y=Jf(a\hat{y})$, and the next nearest neighbors, $J_d=Jf(a\hat{x}+a\hat{y})$ and $J'_d=Jf(a\hat{x}-a\hat{y})$ [Fig. 1(b)]. Their relative magnitudes and signs depend sensitively on the dipole tilting angle $\theta$ and $\phi$. One example is shown in Fig. 1(b) for fixed $\phi=25^\circ$. At small $\theta$, $J_x\sim J_y$ dominates because it is about three times that of $J_d\sim J'_d$. The situation is reminiscent of the $J_1$-$J_2$ model in the regime of the N$\acute{\textrm{e}}$el order.  As $\theta$ is increased, $J_d$ and $J'_d$ grow relative to $J_x$ and $J_y$. The system becomes more frustrated due to the increased competition of the exchanges. This is the most interesting parameter region. Around $\theta\simeq 40^\circ$, $J_x$ and $J_d$ vanish while $J'_d\sim 0.4 J_y$. The model can be viewed as coupled Heisenberg chains. For even larger $\theta$, $J_x$ and $J_d$ switch signs to become ferromagnetic, and the stripe order is expected. Clearly, the physics of $H_{d}$ is much richer than the $J_1$-$J_2$ model. {\it In fact, the two models only overlap at one single point}, $\theta=\phi=0$, where $J_2/J_1=1/2\sqrt{2}\approx 0.35$ and the system is N$\acute{\textrm{e}}$el ordered. 

The degree of frustration can be measured by the ``spin gap" $\Delta$, the energy difference between the ground and the first excited state, from exact diagonalization of $H_d$ for a $4\times 4$ lattice~\cite{noteS}. For example, we observe a pronounced peak in $\Delta$ around $\theta\sim 28^\circ$ for $\phi=25^\circ$, which indicates strong frustration and points to a gapped, spin disordered ground state \cite{refId0}. For fixed $\phi=35^\circ$, the spin structure factor shows a clear peak at $(\pi,\pi)$ for $\theta\sim 15^\circ$ for the N$\acute{\textrm{e}}$el order, a peak at $(0,\pi)$ for $\theta\sim 50^\circ$ for the stripe order, but no well defined peaks around $\theta\sim 35^\circ$, consistent with the argument above.

\textit{Spin-wave and Schwinger-boson theory.}---First, we obtain a coarse phase diagram of $H_d$ on the $(\theta,\phi)$ plane using two widely adopted analytical methods in frustrated quantum magnetism. This will help identify the interesting regions for the more expensive tensor network calculations to focus on. The starting point is the classical solution of $H_d$ by the Luttinger-Tisza method~\cite{LuttingerTisza}. $H_d$ is of the form $\sum_{ij}J_{ij}\mathbf{S}_i\cdot\mathbf{S}_j$ with hard spin constraint $\mathbf{S}_i=S$ and $J_{ij}$ only depends on $\mathbf{r}_i-\mathbf{r}_j$. A theorem states that the classical ground state is a planar spin spiral, $\Bs_\Br/S=\hat{x}\cos(\Bq\cdot\Br)+\hat{y}\sin(\Bq\cdot\Br)$ with an ordering wave vector $\Bq = (Q_x,Q_y)$ \cite{kaplan2007spin}. The classical phase diagram~\cite{noteS} consists of three phases. The first is the N$\acute{\textrm{e}}$el order corresponding to $\Bq=(\pi,\pi)$ for small $\theta$. The second is the stripe phase with $\Bq=(0,\pi)$ for large $\theta$ but not too large $\phi$. These two spin orders are collinear. The third, spiral phase fills the rest of the phase diagram, for large $\theta$ and $\phi$, where $\Bq$ varies continuously and, in general, is incommensurate with the lattice.

Beyond the classical limit, quantum fluctuations will suppress the magnetic order and shift the phase boundary. These effects can be described qualitatively by modified spin wave theory ~\cite{MSW,SelfCSW,SpinW1}.  In the Holstein-Primakoff representation, we expand $H_d$ in a series of $1/S$ and keep up to the quartic order of bosonic operators, i.e., we take into account the interactions between the linear spin waves. The bosonic Hamiltonian is solved by self-consistent mean field theory~\cite{noteS}. The result is summarized in Fig. 2(a). We find that the phase boundary of the N$\acute{\textrm{e}}$el (stripe) phase moves towards smaller (larger) $\theta$ values, opening up an intermediate region in between where the magnetization vanishes. The spiral phase also recedes to higher $\phi$ values. We label this quantum paramagnetic region with QP. This is precisely the region where the various exchanges compete and the system is most frustrated. 

Alternatively, we can take into account quantum fluctuations by the rotationally invariant Schwinger boson mean field theory which is nonperturbative in $S$ \cite{PhysRevB.38.316,PhysRevLett.66.1773}. It is a well tested method capable of describing both magnetically ordered and spin liquid states of frustrated spin models~\cite{mezio2013broken,merino2014spin,lee2014emergent,SchwingerJ1J2}. 
The resulting phase diagram is shown in Fig. 2(b). Here, each magnetic order corresponds to condensation of bosons at a certain wave vector $\Bq$. Within a finite strip region labeled by QP between the N$\acute{\textrm{e}}$el and stripe phase, the condensation fraction vanishes and the spin excitations are gapped, corresponding to a quantum paramagnetic phase. The fact that two different approximations agree on the existence of QP indicates that it must be a robust feature of the model $H_d$.

\begin{figure}
\includegraphics[width=0.5\textwidth]{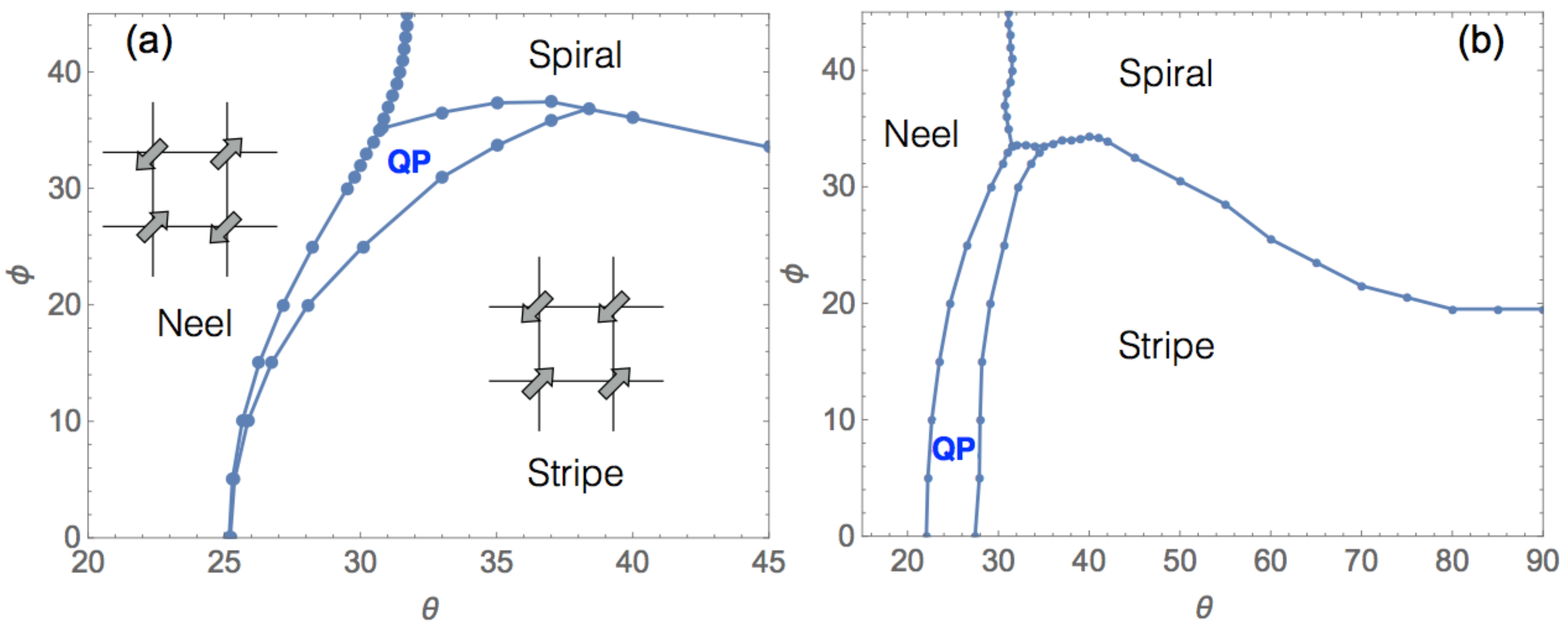}
\caption{Phase diagram of $H_d$ from (a) modified spin wave theory and (b) Schwinger boson mean field analysis. Both methods reveal a QP phase amidst the three long ranged ordered phases: N$\acute{\textrm{e}}$el, stripe, and spiral.}
\label{fig:SBphased}
\end{figure} 

\textit{Phase diagram from a tensor network ansatz.}---A variational ansatz based on tensor network states \cite{PEPS1,iPEPS,iPEPS2} has recently emerged
as an accurate and unbiased algorithm for solving two dimensional frustrated quantum spin models~\cite{tnj1j2,compass120,HKagome1,HKagome2}. In this approach, the ground state many-body wave function $|\Psi\rangle$ is constructed from a network of tensors $T_i$ defined on lattice site $i$: $|\Psi\rangle=\tr\prod_i T_i$, where $\tr$ stands for contraction of neighboring tensors. Each tensor $T_i$ has four virtual legs (indices), each with bond dimension $D$ designed to build up the quantum entanglement between lattice sites, and one physical leg representing the spin. We choose a $L\times L$ cluster as the unit cell with periodic boundary conditions. The algorithm starts with $L^2$ random tensors, and imaginary time evolution is used to update the local tensors, $|\psi'\rangle=\exp(-\tau H)|\psi\rangle$, until convergence is achieved. We adopt the simple update scheme~\cite{simpleup} based on singular value decomposition. By using the Trotter-Suzuki formula $\exp(-\tau H)\approx\prod_{i=1}^4\exp(-\tau H_i)+O(\tau^2)$, each iteration of projection for one plaquette can be done using $\exp(-\tau H_i)$ $(i=1,2,3,4)$ in four separate steps, in which each step evolves three sites (a right triangle) in one plaquette with $H_i$ contains only three terms of the Hamiltonian. For example, $H_{1,2}$ contains $J_x$, $J_y$, and $J_d$ terms and $H_{3,4}$ contains $J_x$, $J_y$ and $J'_d$ terms (See Refs. \cite{nnnsimple,tnSSmodel, tnj1j2,noteS}).

The expectation value of a local operator $O_j$ at site $j$, $\langle O_j\rangle=\langle\Psi|O_j|\Psi\rangle/\langle\Psi|\Psi\rangle$, can be computed by tensor contraction, $\langle O_j\rangle=\tr (\mathcal{O}_j\prod_{i\neq j} \mathcal{T}_i)/\tr\prod_i \mathcal{T}_i$ where $\mathcal{T}_i=T_i^\dagger T_i$ and $\mathcal{O}_j=T_j^\dagger O_j T_j$. We evaluate it using an iterative, real space coarse-graining procedure known as the tensor renormalization group which enables one to reach the thermodynamic limit~\cite{LevinTRG,HOTRG1}. In this way, we calculate the order parameters such as magnetization $M=\sqrt{\langle S_x\rangle^2+\langle S_y\rangle^2+\langle S_z\rangle^2}$~\cite{noteS}.

\begin{figure}
\includegraphics[width=0.4\textwidth]{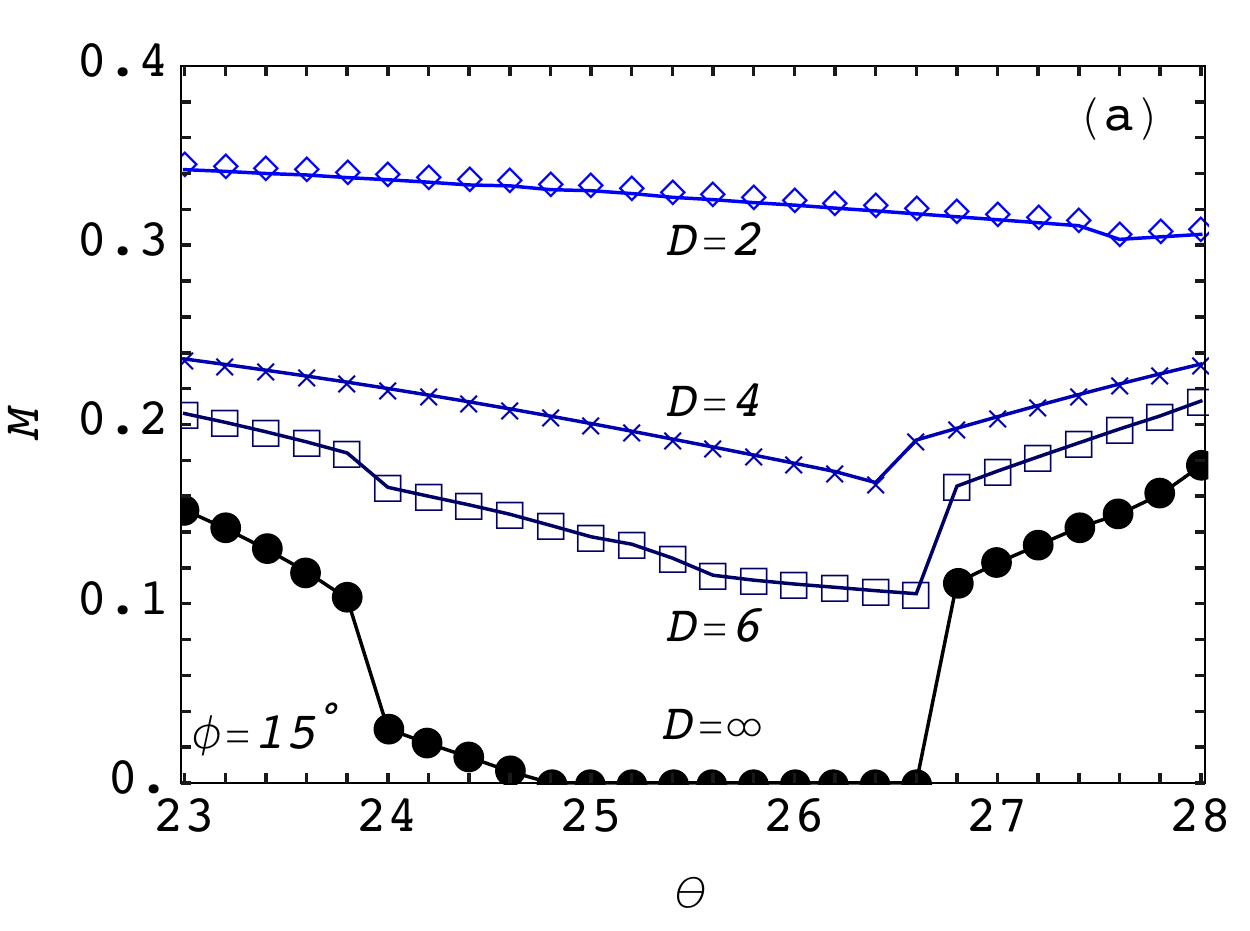}
\includegraphics[width=0.4\textwidth]{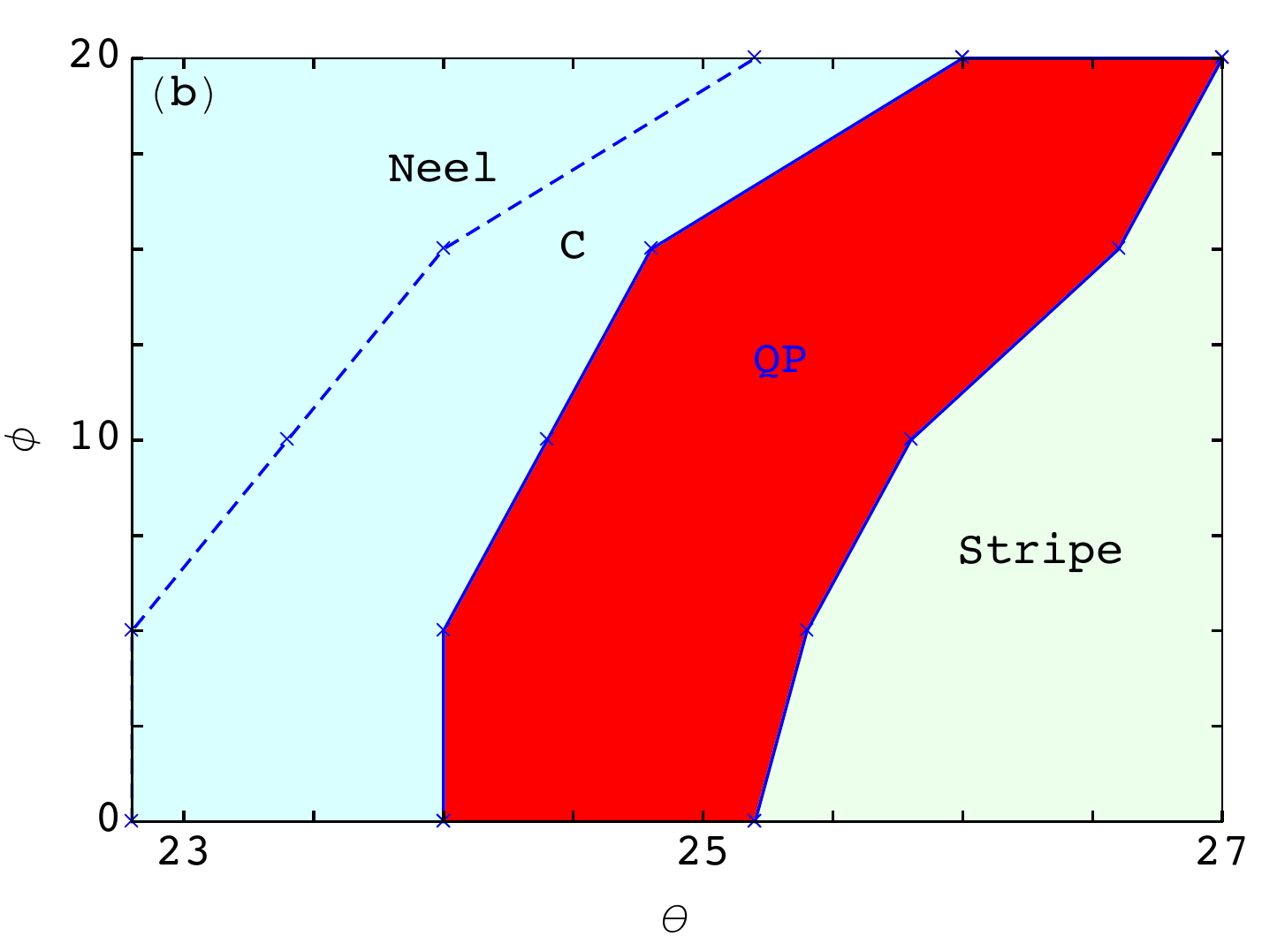}
\caption{ (a) The magnetizations $M$ as functions of $\theta$ for fixed $\phi=15^\circ$ and increasing  $D=2,4,6$. Extrapolation to infinite $D$ by fitting $M$ in polynomials of $1/D$ shows that the magnetic order parameters are suppressed in a finite region of $\theta$, indicating a quantum paramagnetic phase. At $\theta=24.0^\circ$, a sudden drop of $M$ occurs inside the N$\acute{\textrm{e}}$el phase. (b) Phase diagram of $H_d$ for $\phi\le 20^\circ$ obtained from the tensor network ansatz showing a spin-disordered, QP phase sandwiched between the N$\acute{\textrm{e}}$el and stripe phases, broadly consistent with Fig. 2. Region C still has N$\acute{\textrm{e}}$el order, the dashed line indicates where the magnetization $M$ drops suddenly. }
\label{fig:trgpd}
\end{figure} 

With increasing $D$, quantum fluctuations beyond spin wave or Schwinger boson analysis are taken into account. The suppression of $M$ is illustrated in Fig. 3(a) for different $D$ values at fixed $\phi=15^\circ$. By extrapolating the results to infinite $D$, we can determine the phase boundary of the N$\acute{\textrm{e}}$el and stripe phases. Repeating the procedure for different $\phi$ values, we obtain the phase diagram Fig. 3(b). It firmly establishes the existence of a finite quantum paramagnetic region (in red), about one degree wide in $\theta$ and persisting from $\phi=0$ up to $\phi=20^\circ$, where the magnetization is completely suppressed to zero. The paramagnetic phase is narrower than the prediction of the Schwinger boson mean field theory which tends to overestimate the spin disordered region. Inside the N$\acute{\textrm{e}}$el phase, there is a sudden drop of $M$.
Note that the spiral phase, in general, is incompatible with the $L\times L$ cluster choice, even for large $L$. So we refrain from carrying out the tensor network ansatz beyond $\phi=20^\circ$. On the other hand, our numerics indicates that the phase boundary presented in Fig. 3(b) is not expected to depend sensitively on $L$ as it varies~\cite{noteS}. Finally, we point out that the quantum paramagnetic phase is a robust feature of the dipolar $XXZ$ model. It persists when $\eta$ is tuned away from the Heisenberg limit, e.g., down to $\eta=0.5$~\cite{noteS}.

It is challenging to pin down the precise nature of the paramagnetic phase found here in the dipolar Heisenberg model. Similar difficulties also arise for the $J_1$-$J_2$ model where the latest DMRG result~\cite{J1J2Plaq} suggests that the paramagnetic region may consist of a subregion with a plaquette valence bond solid (VBS) order and a second, spin liquid or quantum critical region. Possible spin liquid states for the $J_1$-$J_2$ model on square lattices have been classified within the framework of the Schwinger boson mean field theory~\cite{SchwingerJ1J2}. Yet it remains unclear which one is realized in the ground state. It is possible that the QP region of $H_d$ may contain some VBS order. Unlike the $J_1$-$J_2$ model, the C$_4$ rotation symmetry is broken in $H_d$ as soon as the dipoles are tilted, which may disfavor the plaquette VBS. Because of the limitation of the cluster size, we could not accurately compute the dimer correlation functions.
Future numerical work with larger $L$ and $D$ is required to shed light on this open issue.
The new formulation of symmetric tensor networks~\cite{SymmTN,HKagomeZ2SL} and Lanczos iteration~\cite{LanczosTN} seems promising to detect the possible topological order and accessing the excitation spectrum.

In summary, we presented consistent evidence that a quantum paramagnetic phase emerges from the simple physical system of interacting, tilted dipoles confined on square optical lattices. Our analysis of the dipolar Heisenberg model for general $(\theta,\phi)$ adds a new dimension to frustrated quantum magnetism. It allows the exploration of potential spin liquids beyond the $J_1$-$J_2$ model which has not been realized cleanly so far. For KRb, $J$ is about 100 Hz, or 5 nK, similar to the superexchange scale $t^2/U$ of the Fermi Hubbard model recently studied using quantum gas microscope \cite{Parsons1253,Cheuk1260,Parsons2016,Boll1257,Cheuk2016}. Thus, it seems possible to probe the spin order or spin correlations of $H_d$ and related models in future experiments.

 \acknowledgments
We thank Ying-Jer Kao, Bo Liu, Jaime Merino, and Ling Wang for helpful discussions. This work is supported by the U.S. AFOSR Grant No. FA9550-16-1-0006 (H.Z., E.Z., and W.V.L.), NSF PHY-1205504 (E.Z.), and ARO Grant No. W911NF-11-1-0230, the Chinese National Science Foundation through the Overseas Scholar Collaborative Program (Grant No. 11429402) sponsored by Peking University and another grant (No.11227803), and the Strategic Priority Research Program of the Chinese Academy of Sciences (Grant No. XDB21010100) (W.V.L.). Part of the simulations were done at the supercomputing resources provided by the University of Pittsburgh Center for Simulation and Modeling.

%

\pagebreak
\widetext
\begin{center}
\textbf{\large Supplemental Materials for ``Frustrated magnetism of dipolar molecules on a square optical lattice: Prediction of a quantum paramagnetic ground state"}
\end{center}
\begin{center}
\textbf{Haiyuan Zou, Erhai Zhao and W. Vincent Liu}
\end{center}

\setcounter{equation}{0}
\setcounter{figure}{0}
\setcounter{table}{0}
\setcounter{page}{1}
\makeatletter
\renewcommand{\thefigure}{S\arabic{figure}}
\renewcommand{\thetable}{S\arabic{table}}
\renewcommand{\theequation}{S\arabic{equation}}
\renewcommand{\bibnumfmt}[1]{[S#1]}
\renewcommand{\citenumfont}[1]{S#1}
\makeatother

\section{Classical Phase Diagram}
\begin{figure}[h]
\includegraphics[width=0.48\textwidth]{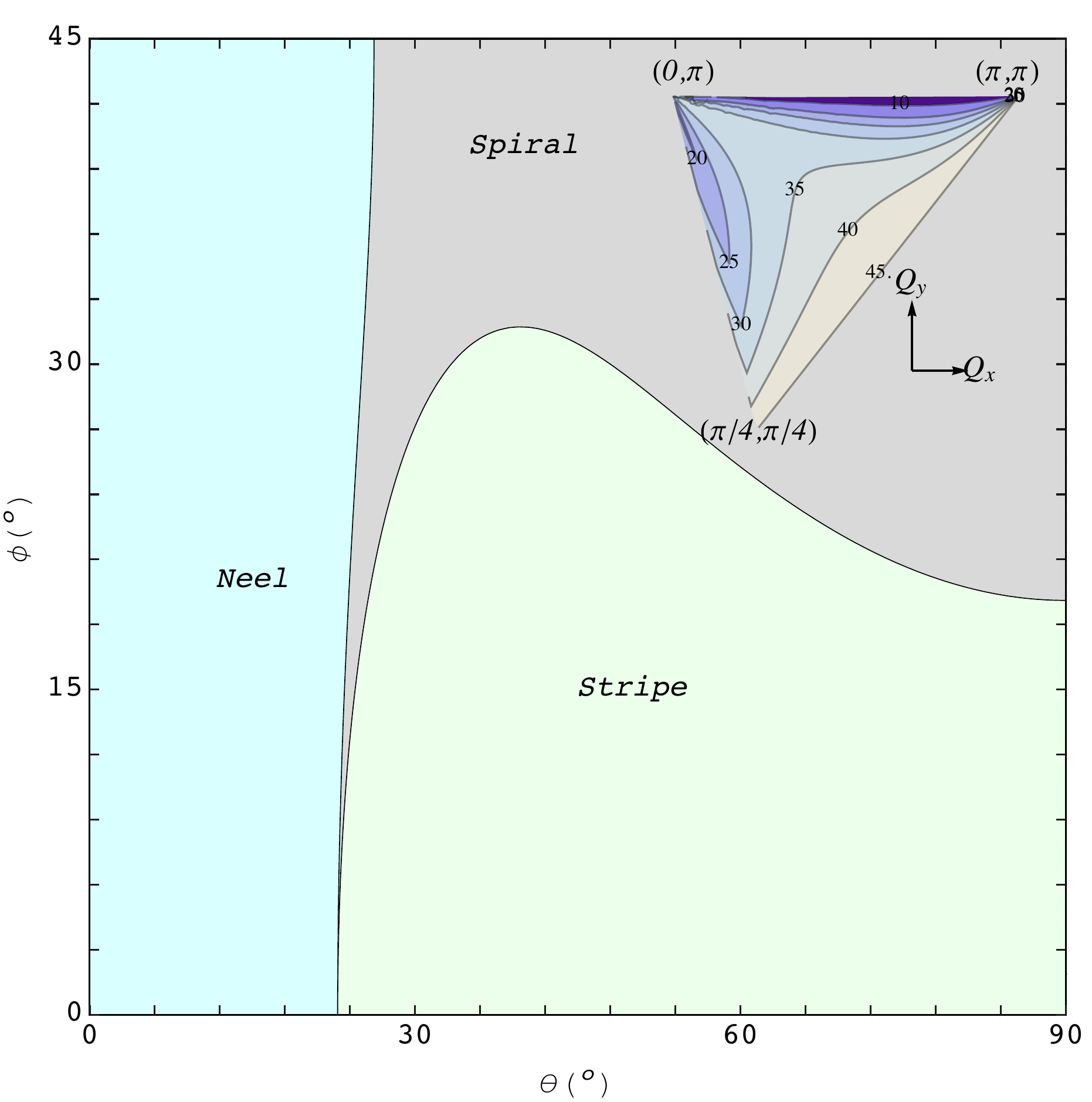}
\caption{The classical phase diagram for the dipolar Heisenberg model on the square lattice. As the dipole tilting angles $\theta$ and $\phi$ are varied, three different phases are realized: the Neel, stripe, and spiral phase. The inset shows the contour plot of the varying wavevector $\Bq$. Each contour corresponds to a horizontal scan at fixed $\phi$ starting from the Neel phase with $\Bq_n=(\pi,\pi)$ going to the right (increasing $\theta$).}
\label{classicpd}
\end{figure}

 The classical ground state of a translationally invariant spin model $\sum_{ij}J_{ij}\Bs_i\cdot \Bs_j$
 on a Bravais lattice can be obtained by minimizing the energy within the planar helix ansatz
\be
\Bs_\Br/S=\hat{x}\cos(\Bq\cdot\Br)+\hat{y}\sin(\Bq\cdot\Br),
\ee
where $\hat{x}$ and $\hat{y}$ form an orthonormal basis and $\Bq = (Q_x,Q_y)$ is the ordering wavevector~\cite{LuttingerTisza2}.
This variational ansatz satisfies the hard spin constraint $|\Bs_\Br|=S$. The classical energy of the dipolar 
Heisenberg model depends on $\Bq$ via
\bea
\label{eq:clH}
\nonumber
2\mathscr{H}_{\textrm{cl}}/{N S^2}&=&J_x\cos(Q_x)+J_y\cos(Q_y)\\ 
&+&J_{d'}\cos(Q_x+Q_y)+J_d\cos(Q_x-Q_y).
\eea
 The set of wavevectors $\Bq$ minimizing the classical energy will be denoted as $\{\Bq\}$.  
 We compare the energies of the incommensurate spiral $\{\Bq_I\}$, the stripe $\Bq_s = (0, \pi)$ and the N$\acute{\textrm{e}}$el $\Bq_n = (\pi, \pi)$ order. The result is the classical phase diagram shown in Fig.~\ref{classicpd}.
Upon crossing the phase boundary, e.g., from the Neel (or the stripe) phase to the incommensurate spiral phase, the wavevector $Q$ varies continuously. 
For example, in the special case of $\phi=45^\circ$ (see the inset of Fig.~\ref{classicpd}), $Q_x=Q_y\equiv Q$, where $Q$ changes continuously from $\pi$ on the phase boundary between Neel and spiral phase to $\pi/4$ at the upper right corner of the $(\theta, \phi)$ diagram.

\section{Exact Diagonalization}
\begin{figure}[h]
\includegraphics[width=0.49\textwidth]{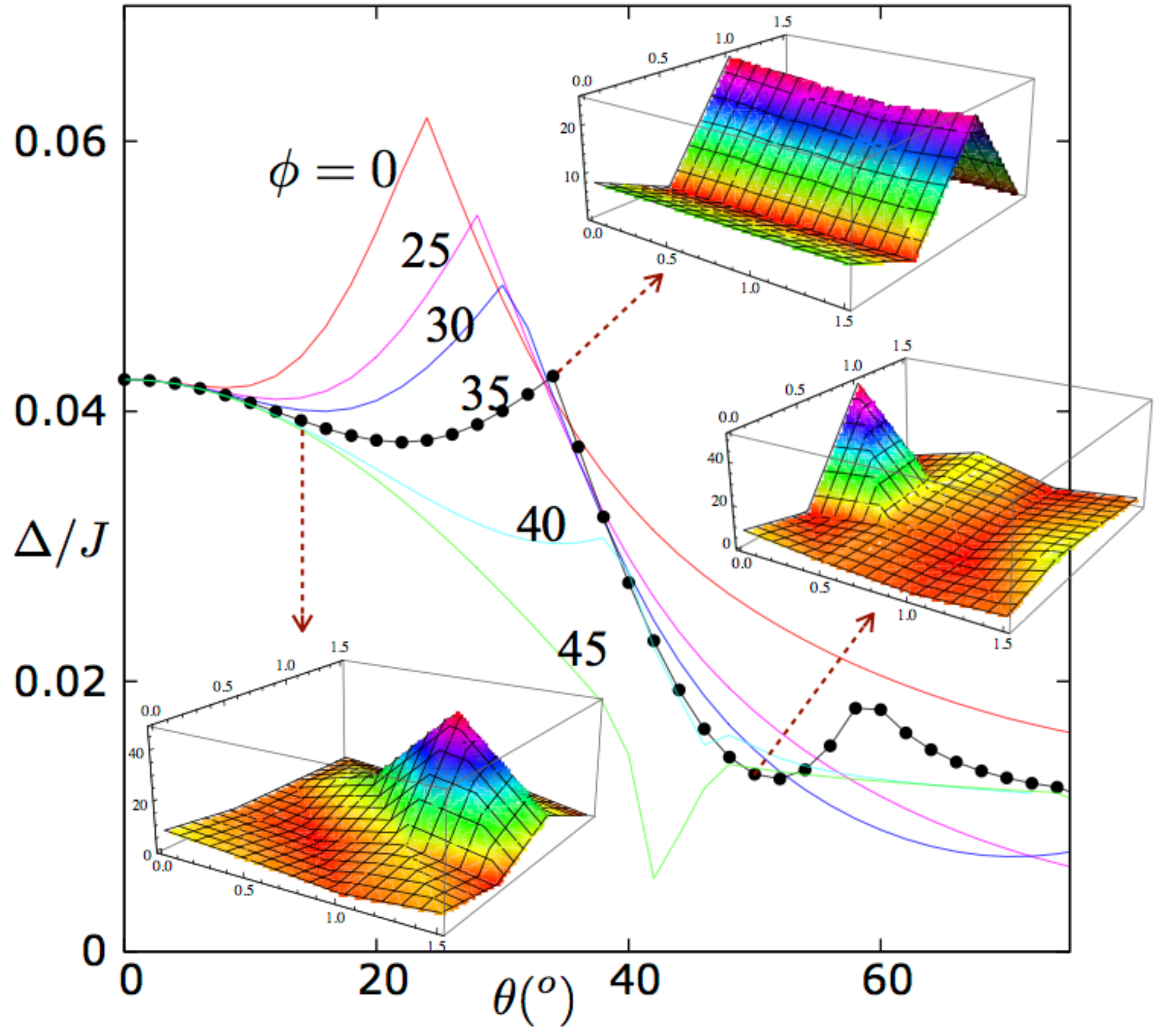}
\caption{The spin gap $\Delta$ as functions of $\theta$ with $\phi=0^\circ,25^\circ,30^\circ,35^\circ,40^\circ, 45^\circ$ are shown. Examples of spin structure factor for Neel order, stripe order, and the transition point between them are shown for $\theta=14^\circ, 50^\circ, 34^\circ$ correspondingly for $\phi=35^\circ$.}
\label{gaped}
\end{figure}

We calculate the ``spin gap" $\Delta$, the energy defference between the ground and the first excited state from exact diagonalization of the dipolar Heisenberg Hamiltonian for a $4\times 4$ lattice. Fig.~\ref{gaped} shows $\Delta$ as functions of $\theta$ for different $\phi$. For $\phi\le 30^\circ$, the pronounced peak at each line indicates strong frustration. For $\phi=35^\circ$, The appearance of the second peak corresponds to the transition from the stripe phase to the incommensurate spiral phase as $\theta$ increases. For $\phi\ge 40^\circ$, the disappearance of the first peak indicates the transition from the Neel phase to the spiral phase. 

\section{Schwinger Boson Theory}
We outline the Schwinger boson mean field theory (SBMF) of the dipolar Heisenberg model. The starting point is the bosonic representation of the spin operators
\bea
\nonumber
S^{+}=a^\dagger b,\\
S^{-}=b^\dagger a,\\ \nonumber
S_z=\frac{1}{2}(a^\dagger a- b^\dagger b).
\eea
with the constraint
\be
\frac{1}{2}(a^\dagger a + b^\dagger b)=S. \label{cons}
\ee
For the square lattice, introduce the antiferromagnetic ($A$) and ferromagnetic ($B$) bond operators
\bea
A_{ij}=\frac{1}{2}(a_i b_j - b_i a_j),\\
B_{ij}=\frac{1}{2}(a^\dagger_i a_j + b^\dagger_i b_j).
\eea 
In terms of the bond operators, the spin exchange term becomes
\be
\Bs_i \cdot \Bs_j=:B^\dagger_{ij} B_{ij}:-A^\dagger_{ij} A_{ij},
\ee
where $::$ means normal order of bosonic operators. Note that $B$ and $A$ are related by operator identity $:B^\dagger_{ij} B_{ij}:+A^\dagger_{ij} A_{ij}=S^2$. 
We adopt the rotational invariant formulation of SBMF and perform mean field decoupling for both $A$ and $B$,
\be
\label{eq:sds}
\Bs_i \cdot \Bs_j\simeq [\beta_{ij} B_{ij}-\alpha_{ij}A_{ij} + h.c. ]-|\beta_{ij}|^2+|\alpha_{ij}|^2,
 \ee
 where
 \bea
 \alpha_{ij}=\langle A^\dagger_{ij}\rangle,\\
 \beta_{ij}=\langle B^\dagger_{ij}\rangle,
 \eea
and $\langle ... \rangle$ denotes the ground state expectation value. This is known to perform better in describing the phases of frustrated spin systems compared to antatz that only keep either $A$ or $B$. Also, within SBMF, the constraint Eq. \eqref{cons} is only enforced on average by introducing the Lagrange multiplier $\lambda$. The mean field Hamiltonian then takes the form
 \bea
\mathscr{H}_{MF}&=&\frac{1}{2}\sum_{i\neq j}J_{ij} [\beta_{ij} B_{ij}-\alpha_{ij}A_{ij}+ h.c.- |\beta_{ij}|^2+|\alpha_{ij}|^2 ] \nonumber \\
&+& \lambda\sum_{i}[a^\dagger_i a_i +
b^\dagger_i b_i-2S].
\eea

Long range magnetic order corresponds to condensation of the $a$ and/or $b$ bosons. To treat the condensate fraction, we decompose each operator into
\bea
a_i = \tilde{a}_i + x_i,\;\; x_i=\langle a_i \rangle ,\\
b_i = \tilde{b}_i +y_i,\;\; y_i=\langle b_i \rangle ,
\eea
where $x_i$ and $y_i$ are c-numbers describing the condensate, while operators $\tilde{a}_i$ and $\tilde{b}_i$ annihilate excitations over the condensate. We assume $\alpha_{ij}=\alpha_\delta$ with $\boldsymbol{\delta}=\Br_j-\Br_i$ and similarly for $\beta_{ij}$. Namely they only depends on $\boldsymbol{\delta}$ and not on $\Br_i$. For our model, it is sufficient to keep $\boldsymbol{\delta}=\pm \hat{x}$, $\pm \hat{y}$, $\pm\hat{x}\pm\hat{y}$, i.e. the nn and nnn couplings.
Fourier transform to $\mathbf{k}$ space, e.g. $\tilde{a}_i\rightarrow \tilde{a}_k$,  $\mathscr{H}_{MF}$ becomes a quadratic form of operators $\tilde{a}_k$, $\tilde{b}_k$ and c-numbers $x_k$, $y_k$. In accordance with the classical analysis, we assume $x_k$ and $y_k$ are nonzero only at a pair of wave vector $\pm\mathbf{Q}/2$. It is then diagonalized by a standard Bogoliubov transformation, 
  \bea
\mathscr{H}_{MF}&=&\sum_{k} [c^\dagger_k c_k + d^\dagger_k d_k +1]\omega_k +  \frac{N}{2} \sum_\delta J_{\delta} (|\alpha_\delta|^2-|\beta_\delta|^2) \nonumber \\
 &+& \sum_{\mathbf{k}=\pm \mathbf{Q}/2} [\beta_k (|x_k|^2+|y_k|^2) +(ix_ky_{-k}\alpha^*_k + h.c.)]\nonumber \\
 &-&N\lambda (2S +1) +\lambda\sum_{\mathbf{k}=\pm \mathbf{Q}/2} (|x_k|^2+|y_k|^2).
 \eea
Here $N$ is the number of lattice sites, $c_k$ and $d_k$ are the eigenmodes of spin excitations with dispersion
\be
\omega_k=\sqrt{(\lambda+\beta_k)^2-|\alpha_k|^2}.
\ee
and 
\bea
\alpha_k=\frac{1}{2}\sum_\delta J_\delta \sin (\mathbf{k}\cdot\boldsymbol{\delta}) \alpha_\delta,\\
\beta_k=\frac{1}{2}\sum_\delta J_\delta \cos (\mathbf{k}\cdot\boldsymbol{\delta}) \beta_\delta.
\eea
We adopt the sprial ansatz $x_i=\sqrt{2m}\cos(\frac{\mathbf{Q}}{2}\cdot\Br_i)$, $y_i=\sqrt{2m}\sin(\frac{\mathbf{Q}}{2}\cdot\Br_i)$. Then $x_{Q/2}=\sqrt{Nm/2}$, $y_{Q/2}=-ix_{Q/2}$. 

Minimizing the SBMF ground energy with respect to the variational parameters $\{\lambda, \alpha_\delta, \beta_\delta,x_\frac{\mathbf{Q}}{2},y_\frac{\mathbf{Q}}{2}\}$ leads to the self-consistency equations,
\bea
S+\frac{1}{2}=\frac{1}{2N}\sum_k \frac{\lambda+\beta_k}{\omega_k}+m, \\
\alpha_\delta=\frac{1}{2N}\sum_k \frac{\alpha_k}{\omega_k}\sin (\mathbf{k}\cdot\boldsymbol{\delta})+m\sin (\frac{\mathbf{Q}}{2}\cdot\boldsymbol{\delta}),\\
\beta_\delta=\frac{1}{2N}\sum_k \frac{\lambda+\beta_k}{\omega_k}\cos (\mathbf{k}\cdot\boldsymbol{\delta})+m\cos (\frac{\mathbf{Q}}{2}\cdot\boldsymbol{\delta}),\\
\lambda+\beta_\frac{\mathbf{Q}}{2} = \alpha_\frac{\mathbf{Q}}{2}.
\eea
The last equation is equivalent to the requirement that $\mathbf{Q}$ is chosen to be the minimum of $\omega_k$.
And the SBMF ground state energy simplifies to
\be
E_{MF}=\sum_k \omega_k -N\lambda (2S +1) -\frac{N}{2} \sum_\delta J_\delta (|\beta_\delta|^2-|\alpha_\delta|^2).
\ee

In the large $S$ limit, we have $m\simeq S$, $\alpha_\delta= m \sin (\frac{\mathbf{Q}}{2}\cdot\boldsymbol{\delta})$, $\beta_\delta= m \cos (\frac{\mathbf{Q}}{2}\cdot\boldsymbol{\delta})$, $\lambda=-\sum_\delta J_\delta \cos ( \mathbf{Q} \cdot\boldsymbol{\delta})$, and 
\be
E_\mathrm{\textrm{cl}}=\frac{1}{2}NS^2\sum _\delta J_\delta \cos ( \mathbf{Q} \cdot\boldsymbol{\delta}),
\ee
which agrees with the classical result as expected.

\section{Modified Spin Wave Theory}

We represent the spin operator using Holstein-Primakoff (HP) bosons,
\bea
\nonumber
S^{-}&=&a^\dagger \sqrt{2S-a^\dagger a},\\
S^{+}&=&\sqrt{2S-a^\dagger a} a,\\ \nonumber
S_z&=&S-a^\dagger a.
\eea
Proper number of boson operators are introduced for the two-sublattice case (Neel phase) and the four-sublattice case (stripe phase). Take the two-sublattice for example, $a_i$ (or $b_j$) are Bose annihilation operators on the A (or B) sublattice. 
The dipolar Heisenberg Hamiltonian can then be expanded in series of boson operators, 
\be
\mathscr{H}_{SW}=\mathscr{H}_{\textrm{cl}}+\mathscr{H}^{(2)}+\mathscr{H}^{(4)},
\ee
where the classical part $\mathscr{H}_{\textrm{cl}}$ is given previously in Eq.~\ref{eq:clH}, the quadratic part $\mathscr{H}^{(2)}$ is
\bea
\mathscr{H}^{(2)}/S&=&J_x\sum_{nn_x}(a^\dagger_i a_i+b^\dagger_j b_j-a^\dagger_i b^\dagger_j-a_i b_j) \nonumber \\
&+&J_y\sum_{nn_y}(a^\dagger_i a_i+b^\dagger_j b_j-a^\dagger_i b^\dagger_j-a_i b_j) \nonumber \\
&+&J_{d'}\sum_{nnn_1}(a^\dagger_i a_{i'}+b^\dagger_j b_{j'}-a^\dagger_i a_i-b^\dagger_j b_j) \nonumber \\
&+&J_{d}\sum_{nnn_2}(a^\dagger_i a_{i'}+b^\dagger_j b_{j'}-a^\dagger_i a_i-b^\dagger_j b_j),
\eea
and the quartic part $\mathscr{H}^{(4)}$ is 
\bea
4\mathscr{H}^{(4)}&=&J_x\sum_{nn_x}(a^\dagger_i a_ia_ib_j+a_jb^\dagger_j b_jb_j-2a^\dagger_ia_ib^\dagger_jb_j) \nonumber \\
&+&J_y\sum_{nn_y}(a^\dagger_i a_ia_ib_j+a_jb^\dagger_j b_jb_j-2a^\dagger_ia_ib^\dagger_jb_j) \nonumber \\
&+&J_{d'}\sum_{nnn_1}(a^\dagger_ia^\dagger_{i'}a_ia_{i'}-a^\dagger_i a^\dagger_{i'}a_{i'}a_{i'}) \nonumber \\
&+&J_{d'}\sum_{nnn_1}(b^\dagger_jb^\dagger_{j'}b_jb_{j'}-b^\dagger_j b^\dagger_{j'}b_{j'}b_{j'}) \nonumber \\
&+&J_{d}\sum_{nnn_2}(a^\dagger_ia^\dagger_{i'}a_ia_{i'}-a^\dagger_i a^\dagger_{i'}a_{i'}a_{i'}) \nonumber \\
&+&J_{d}\sum_{nnn_2}(b^\dagger_jb^\dagger_{j'}b_jb_{j'}-b^\dagger_j b^\dagger_{j'}b_{j'}b_{j'})+\textrm{H.C.}.
\eea

For the Neel phase, the expectation values of many operator pairs vanish, e.g., 
\be
\langle a_ia_i\rangle=\langle a_ia_{i'}\rangle=\langle a^\dagger_ib_j\rangle=0.
\ee
We define the following nonzero averages of boson operators describing the quantum fluctuations of spins
\bea
f_0\equiv\langle a^\dagger_ia_i\rangle, \nonumber \\
g_1\equiv\langle a_ib_j\rangle, \nonumber \\
f_2\equiv\langle a^\dagger_ia_j\rangle,
\eea
and apply self-consistent mean field decoupling of the quartic terms in $\mathscr{H}^{(4)}$ 
\bea
\label{eq:quartic}  
a^\dagger_ia_ia_ib_j&=&f_0 a_ib_j+g_1 a^\dagger_ia_i-f_0 g_1, \nonumber \\
a^\dagger_ia^\dagger_{i'}a_{i'}a_{i'}&=&f_0 a^\dagger_ia_{i'}+f_2 a^\dagger_{i'}a_{i'}-f_0 f_2, \nonumber \\
a^\dagger_ia_ib^\dagger_jb_j&=&(1-\lambda_1)(f_0 b^\dagger_jb_j+f_0 a^\dagger_ia_i-f_0^2) \nonumber  \\
&+&\lambda_1(g_1a_ib_j+g_1a^\dagger_ib^\dagger_j-g_1^2), \nonumber \\
a^\dagger_ia_ia^\dagger_{i'}a_{i'}&=&(1-\lambda_2)(f_0 a^\dagger_{i'}a_{i'}+f_0 a^\dagger_ia_i-f_0^2) \nonumber  \\
&+&\lambda_2(f_2a^\dagger_ia_{i'}+f_2a^\dagger_{i'}a_i-f_2^2),
\eea
where $0\le\lambda_1,\lambda_2\le 1$ are parameters determined by minimizing the ground state energy. The magnitudes of $\lambda_1$ or $\lambda_2$ describe the competition between the diagonal and off-diagonal terms of spin deviation operators. 

After Bogoliubov transformation to diagonalize the resulting Hamiltonian, the self-consistent equations can be solved by minimizing the ground state energy $E_0$ with respect to the variational parameters $\{ \lambda_1, \lambda_2\}$. 
The energy and the staggered magnetization are given by
\bea
E_0&=&\mathscr{H}_{\textrm{cl}}+E_1+\sum_{\mathbf{k}}\epsilon_k, \\
\langle S_z\rangle&=&S-f_0,
\eea
where
\bea
&&E_1=\nonumber\\ 
&&J_x[(1-\lambda_1)(f_0^2-f_0)+\lambda_1g_1^2+(\frac{1}{2}-f_0)g_1+S] \nonumber \\
&+&J_y[(1-\lambda_1)(f_0^2-f_0)+\lambda_1g_1^2+(\frac{1}{2}-f_0)g_1+S] \nonumber \\
&-&J_{d'}[(1-\lambda_2)(f_0^2-f_0)+\lambda_2f_2^2+(\frac{1}{2}-f_0)f_2+S] \nonumber \\
&-&J_{d}[(1-\lambda_2)(f_0^2-f_0)+\lambda_2f_2^2+(\frac{1}{2}-f_0)f_2+S], 
\eea
and
\be
\epsilon_k=\sqrt{h_k^2-\Delta_k^2},
\ee
with 
\bea
h_k&=&J_x[S-(1-\lambda_1)f_0+\frac{1}{2}g_1] \nonumber \\
&+&J_y[S-(1-\lambda_1)f_0+\frac{1}{2}g_1] \nonumber \\
&-&J_{d'}[S-(1-\lambda_2)f_0+\frac{1}{2}f_2)] \nonumber \\
&-&J_{d}[S-(1-\lambda_2)f_0+\frac{1}{2}f_2)] \nonumber \\
&+&J_{d'}[\cos(k_x+k_y)(S-\frac{f_0}{2}+\lambda_2f_2)] \nonumber \\
&+&J_d[\cos(k_x-k_y)(S-\frac{f_0}{2}+\lambda_2f_2)],
\eea
and
\bea
\Delta_k&=&J_x\cos(k_x)(\frac{f_0}{2}-S-\lambda g_1) \nonumber \\
&+&J_y\cos(k_y)(\frac{f_0}{2}-S-\lambda g_1).
\eea

The self-consistency equations are
\bea
f_0&=&\frac{1}{N}\sum_{\mathbf{k}}(\frac{h_k}{\epsilon_k}-1), \nonumber \\
g_1&=&-\frac{1}{N}\sum_{\mathbf{k}}\frac{\Delta_k}{2\epsilon_k}(\cos k_x+\cos k_y), \nonumber \\
f_2&=&\frac{1}{N}\sum_{\mathbf{k}}\frac{1}{\epsilon_k}\cos k_x\cos k_y.
\eea
The criterion for Neel order is a finite $\langle S_z\rangle_{Neel}$.  

For the stripe case, a similar procedure can be applied except that four types of boson operators should be introduced. Correspondingly, three variational parameters $\mathbf{\lambda}_i$ are needed due to the difference between $x$ and $y$ directions. Using a similar self-consistent mean-field approximation, the boundary of stripe phase can be determined. The criterion for stripe phase is a finite $\langle S_z\rangle_{stripe}$ and real, positive-definite spin deviation operators.  

The mean field phase diagrams in Fig.~2 of the main text obtained by two different methods give us the same qualitatively picture but  different areas of the spin disordered region. This is not surprising, since different spin representations and mean field decoupling schemes are used. For example, in Eq.~\ref{eq:sds}, the expectation values of bond operators are used in SBMF while the quartic terms in Eq.~\ref{eq:quartic} are described by the variational parameters of ${\lambda_1, \lambda_2}$ for the modified spin wave theory.

\section{Tensor Network Ansatz}
\subsection{Simple Update}
We choose a $L\times L$ unit cell (i.e. $L\times L$ local tensors) with different virtual bond dimension $D=2,4,6$ to form the initial tensor network state $|\Psi\rangle$ and set the time interval $\tau=0.005J^{-1}$ for imaginary time evolution iterations for local tensors $|\psi\rangle$,
\be
|\psi'\rangle=\exp(-\tau H)|\psi\rangle,
\ee
until convergence is achieved. 
\begin{figure}[h]
\includegraphics[width=0.5\textwidth]{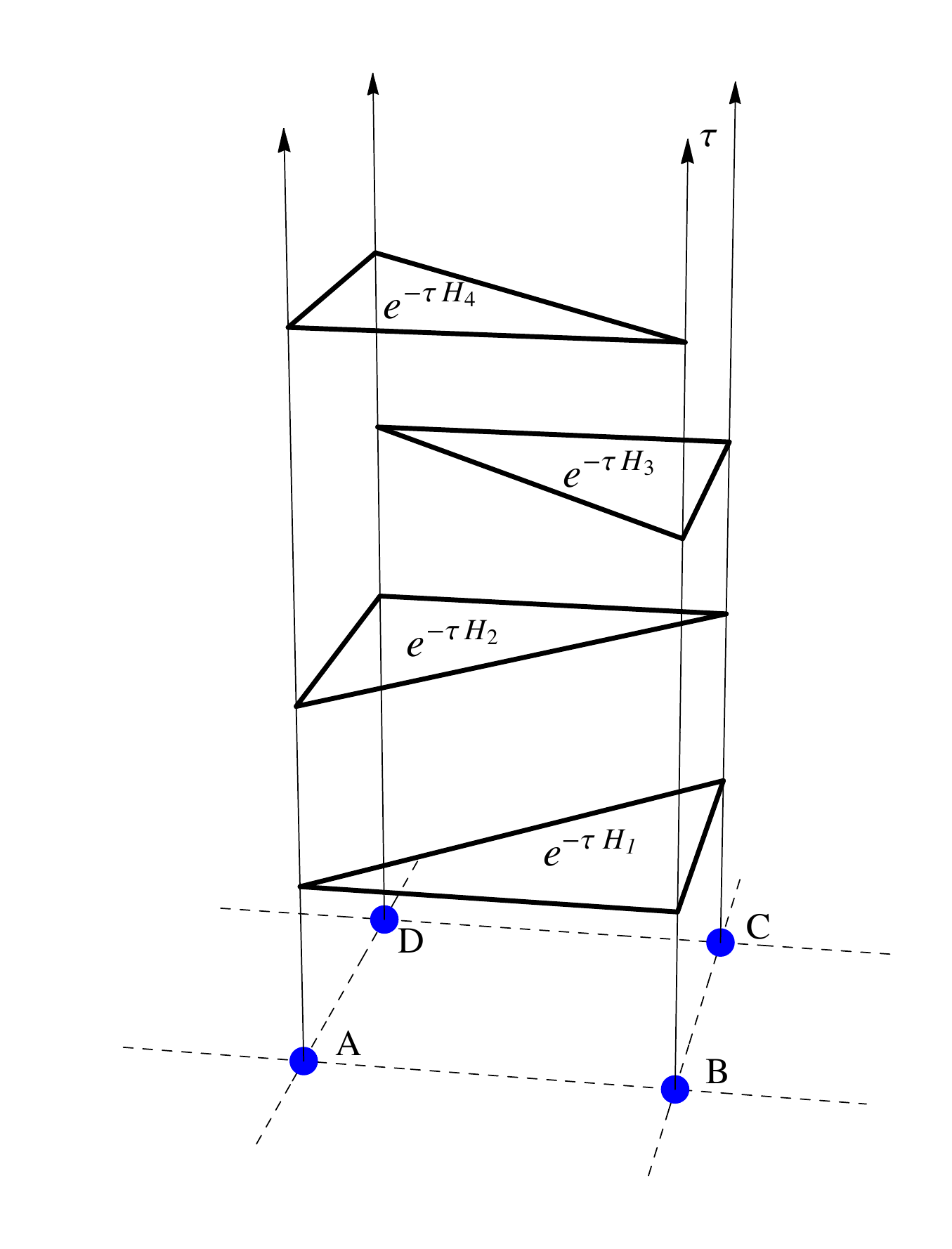}
\caption{The update scheme with a $2\times 2$ unit cell. Tensors $A,B,C,D$ are updated alternately with operators $\exp(-\tau H_i)$ $(i=1,2,3,4)$. In each step, three local tensors are updated.}
\label{fig:step1}
\end{figure} 
Taking $L=2$ as an example and using the Trotter-Suzuki formula~\cite{simpleup2,nnnsimple2}, we can express the projection operator as
\be
\exp(-\tau H)\approx\prod_{i=1}^4\exp(-\tau H_i)+O(\tau^2),
\ee
where
\bea
H_1&=&J_x\mathbf{S}_A \cdot \mathbf{S}_B+J_y\mathbf{S}_B \cdot \mathbf{S}_C+2J_d\mathbf{S}_A \cdot \mathbf{S}_C,\nonumber\\
H_2&=&J_x\mathbf{S}_C \cdot \mathbf{S}_D+J_y\mathbf{S}_A \cdot \mathbf{S}_D+2J_d\mathbf{S}_A \cdot \mathbf{S}_C,\nonumber\\
H_3&=&J_x\mathbf{S}_ C\cdot \mathbf{S}_D+J_y\mathbf{S}_B \cdot \mathbf{S}_C+2J'_d\mathbf{S}_B \cdot \mathbf{S}_D,\nonumber\\
H_4&=&J_x\mathbf{S}_A \cdot \mathbf{S}_B+J_y\mathbf{S}_A \cdot \mathbf{S}_D+2J'_d\mathbf{S}_B \cdot \mathbf{S}_D.
\eea
This means that each iteration of projection can be done using $\exp(-\tau H_i)$ $(i=1,2,3,4)$ in four separate steps for one plaquette. While in each step three out of four tensors are evolved (Fig.~\ref{fig:step1}). 

\subsection{Tensor Renormalization Group}
\begin{figure}[h]
\includegraphics[width=0.5\textwidth]{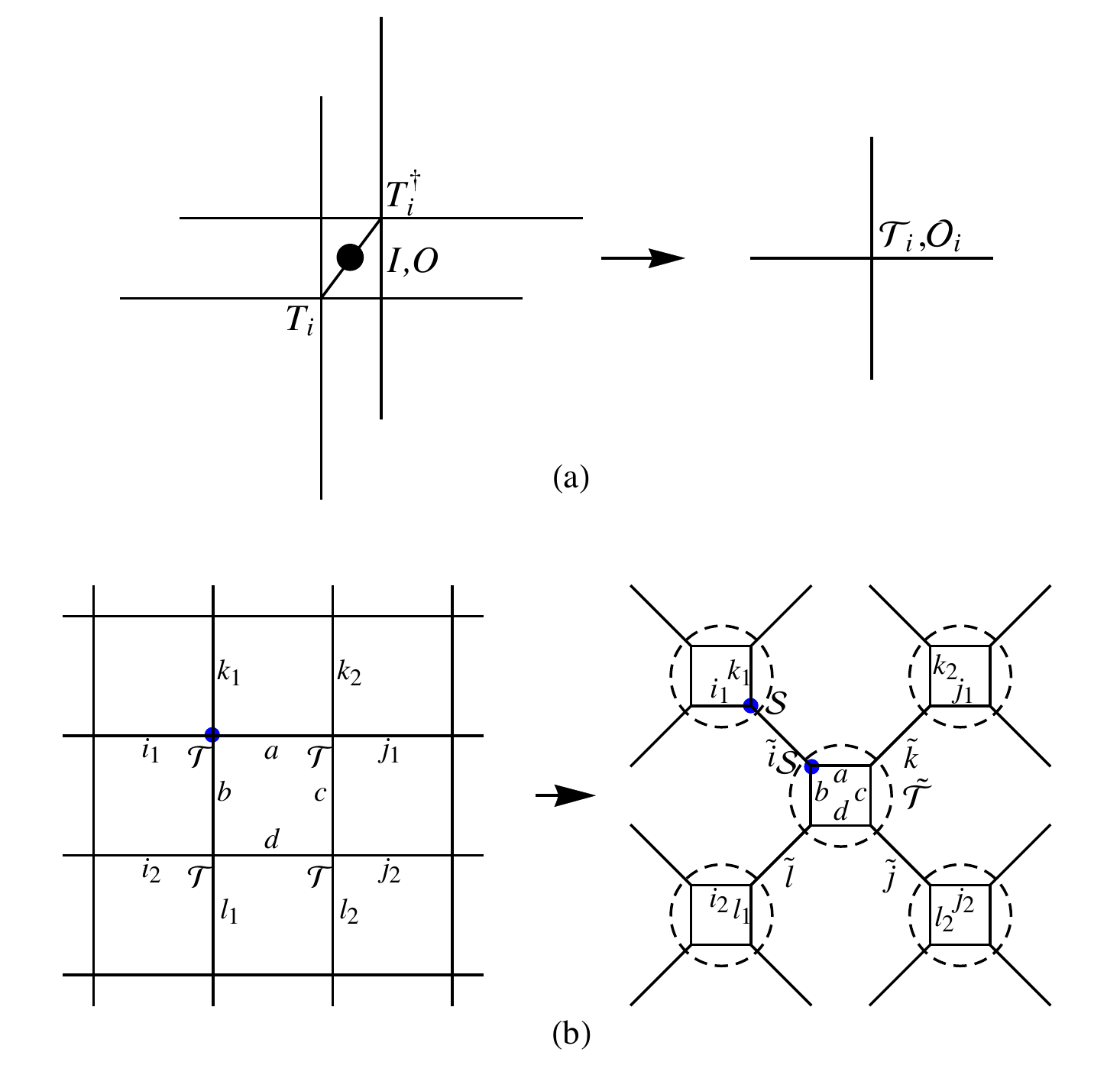}
\caption{TRG steps: (a) New rank-4 local tensors $\mathcal{T}$ (or operating tensors $\mathcal{O}$) are constructed from $T$ and identity $I$ (or $O$); (b) By using singular value decomposition, $\mathcal{T}$ is decompose to two $\mathcal{S}$. Contraction of the inner legs of four $\mathcal{S}$ forms a new $\tilde{\mathcal{T}}$ tensor (dashed circle).}
\label{fig:step2}
\end{figure}

Starting from the converged local tensors $T_i$ obtained from the simple update, one can construct new two-dimensional local tensors (Fig.~\ref{fig:step2}(a)),
\bea
\mathcal{T}_i&=&T_i^\dagger T_i,\nonumber\\
\mathcal{O}_i&=&T_i^\dagger O_i T_i,
\eea
where $O_i$ is an operator.

The expectation value of  $O_i$, $\langle O_i\rangle=\langle\Psi|O_i|\Psi\rangle/\langle\Psi|\Psi\rangle$, can then be obtained by 
\be
\langle O_i\rangle=\tr (\mathcal{O}_i\prod_{j\neq i} \mathcal{T}_j)/\tr\prod_j \mathcal{T}_j,
\ee
in the thermodynamic limit by using Tensor Renormalization Group (TRG) method~\cite{LevinTRG2,HOTRG12}, where tr stands for contraction of neighboring tensors.
Taking the denominator $\tr\prod_i \mathcal{T}_i$ as an example (the numerator can be coarse-grained with the same procedure because the local operator $O$ has the same structure with $\mathcal{T}$). As shown in Fig.~\ref{fig:step2}(b), for each step, one can decompose each $\mathcal{T}$ to two $\mathcal{S}$ via singular value decomposition, 
\be
\mathcal{T}_{i_1,a,k_1,b}\approx\sum_{\tilde{i}}\mathcal{S}_{i_1,k_1,\tilde{i}}\mathcal{S}_{a,b,\tilde{i}}.
\ee
The truncation bond dimension of coarse-graining (dimension of the third leg of $\mathcal{S}$) is set as $\chi$. The new local tensor $\tilde{\mathcal{T}}$ with the same structure as $\mathcal{T}$ can be constructed from contracting the inner legs of four $\mathcal{S}$,
\be
\tilde{\mathcal{T}}_{\tilde{i},\tilde{j},\tilde{k},\tilde{l}}=\sum_{a,b,c,d}\mathcal{S}_{a,b,\tilde{i}}\mathcal{S}_{d,c,\tilde{j}}\mathcal{S}_{a,c,\tilde{k}}\mathcal{S}_{d,b,\tilde{l}}.
\ee
Using $\tilde{\mathcal{T}}$ as the starting tensors, these steps are repeated until $\tr\prod_i \mathcal{T}_i$ is converged. In our TRG calculation, the truncation bond dimension is fixed as $\chi = 8$ to make $D$ the only tuning parameter of the whole procedure.

\subsection{Comparison of Different Unit Cell Sizes}
\begin{figure}[h]
\includegraphics[width=0.45\textwidth]{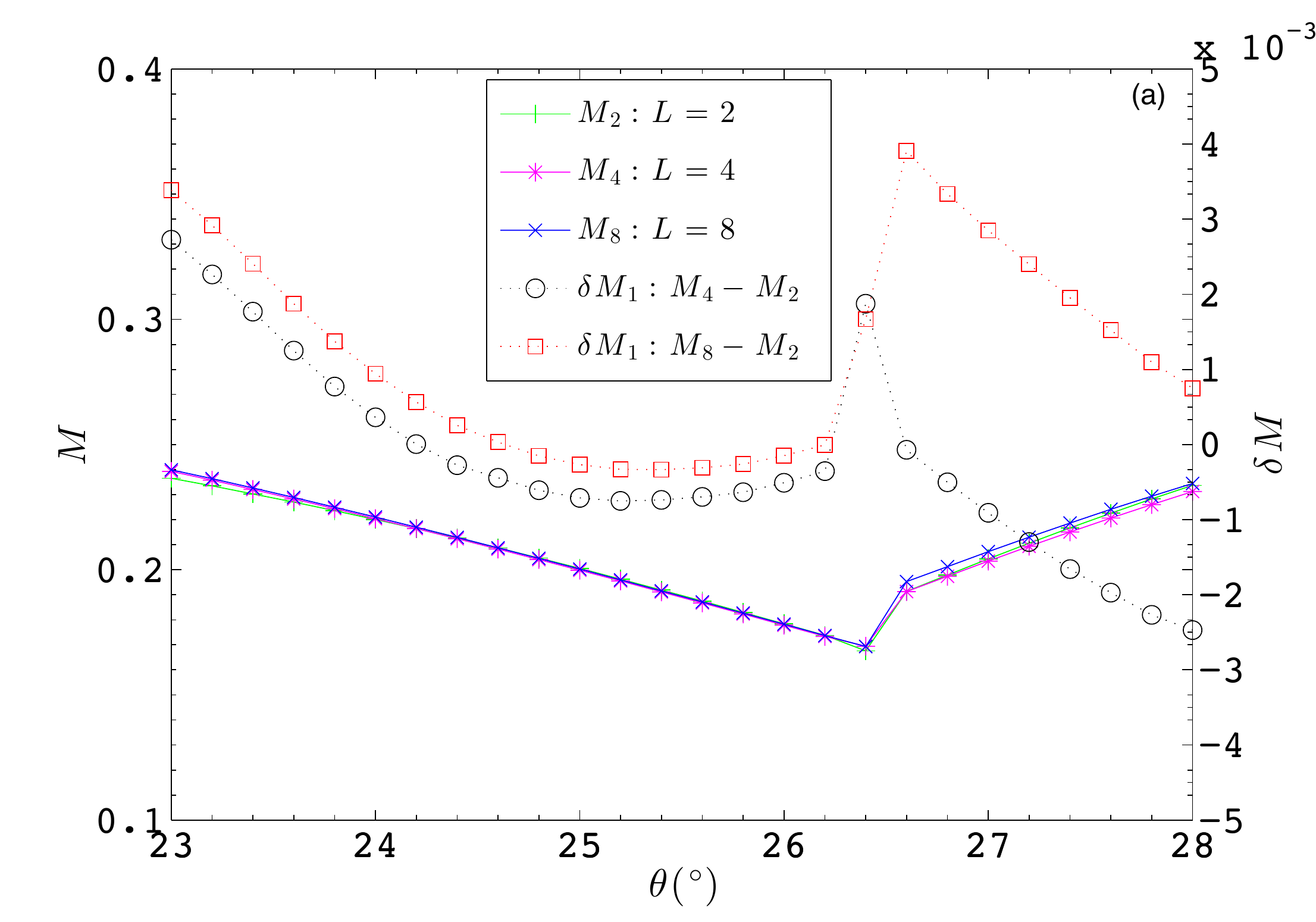}
\includegraphics[width=0.45\textwidth]{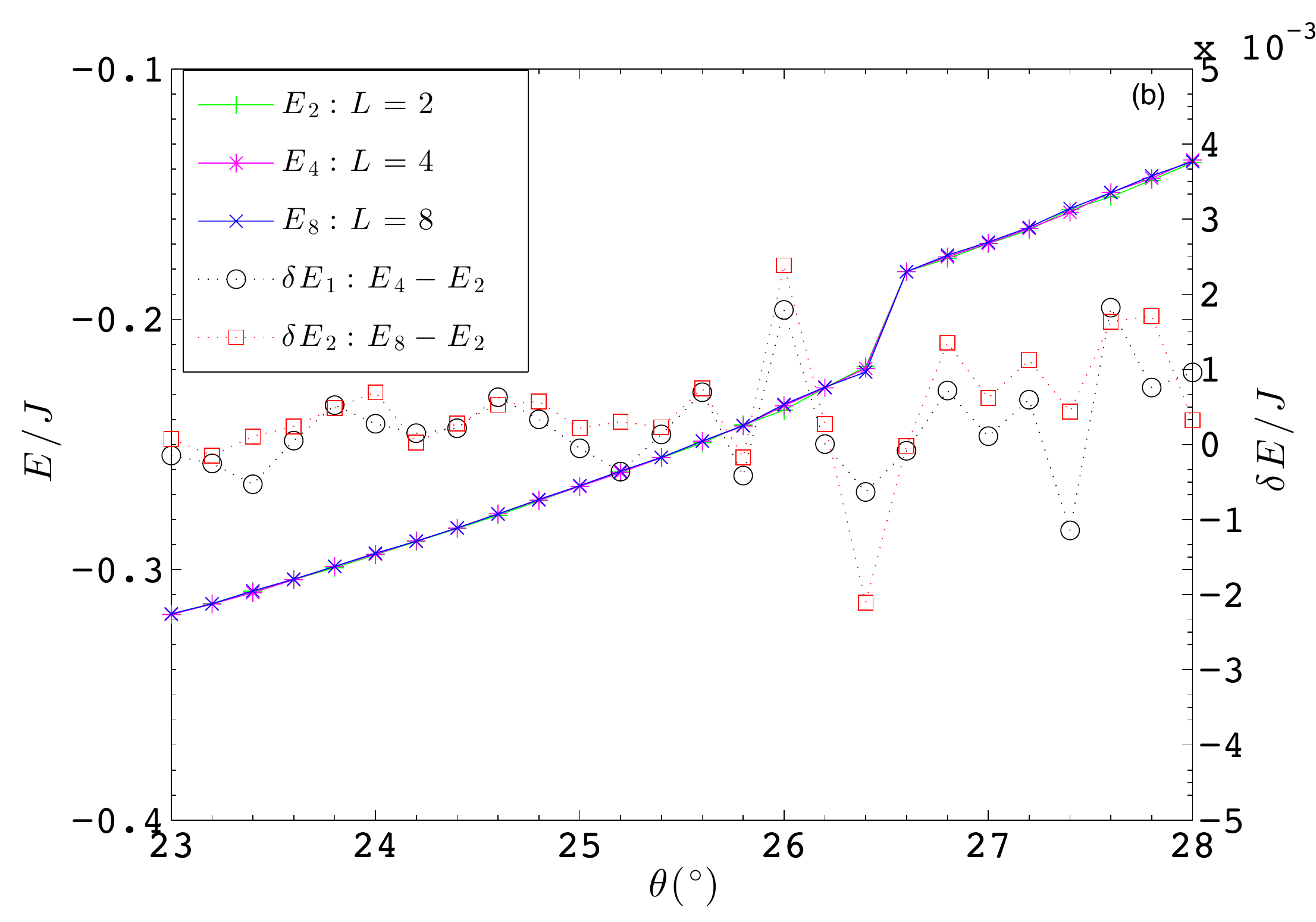}
\caption{(a) Magnetizations $M_{2,4,8}$ and (b) Average energies $E_{2,4,8}$ for $L=2,4,8$ with $D=4$. As $L$ increased,  the differences between magnetizations (or energies) with larger $L$ and those with $L=2$ are very small.}
\label{fig:dEnergy}
\end{figure} 
The phase boundary between the quantum paramagnetic phase and other long-range orders can be inferred from the disappearance of magnetic order parameters. It is crucial to determine whether the phase boundary depends sensitively on $L$, the size of the unit cell. To address this question, we calculate the magnetization $M$ and the average energy $E$ with different unit cell size ($L=2,4,8$) at fixed $D$. Fig.~\ref{fig:dEnergy} shows examples with $D=4$, from which we conclude that increasing $L$ does not increase the accuracy significantly. Thus the scaling of $L$ to larger value gives essentially the same result as $L=2$ and we can use $L=2$ to obtain the phase diagram in the main text. 

\subsection{Average Energy and Extrapolation of $M$}
\begin{figure}[h]
\includegraphics[width=0.42\textwidth]{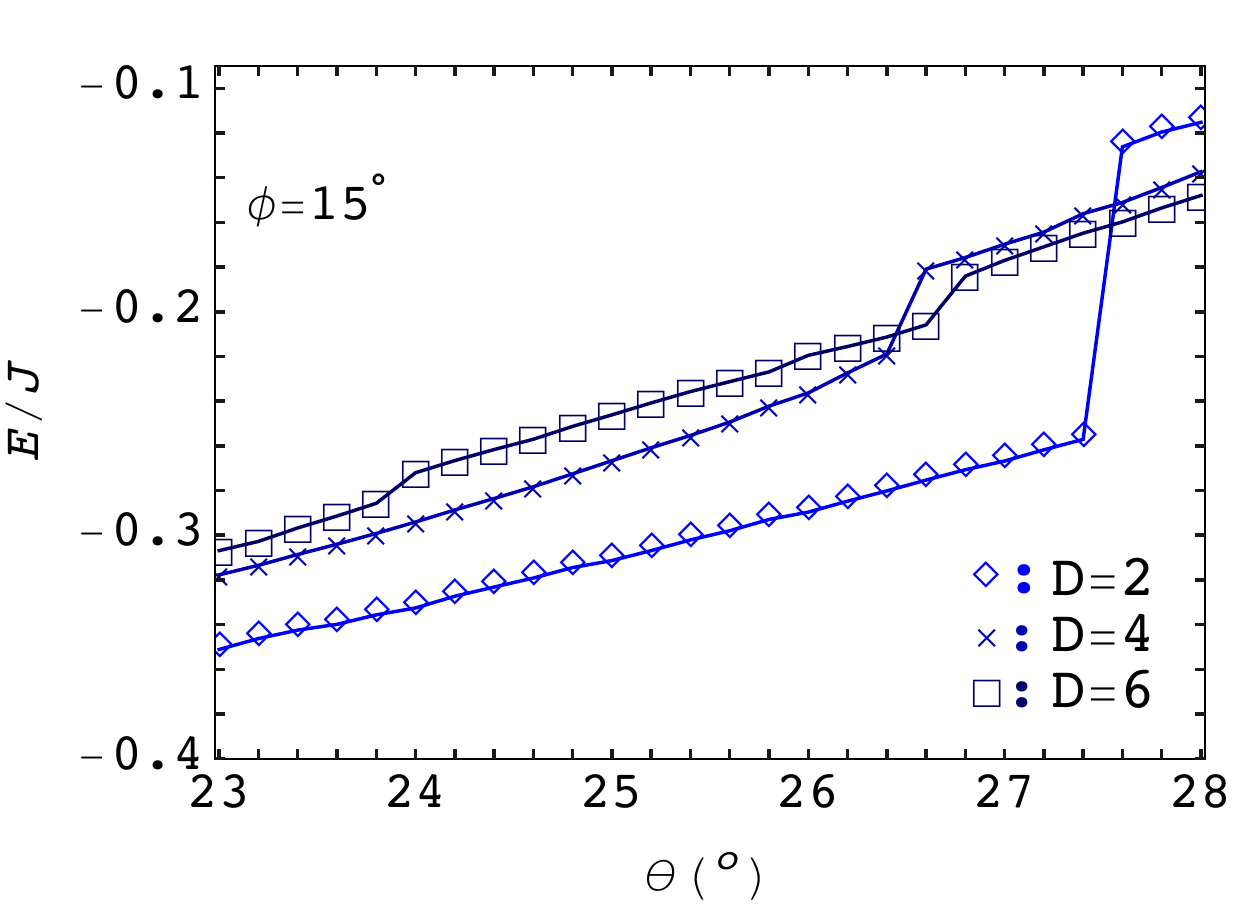}
\caption{ Average energy $E$ as function of $\theta$ for $\phi=15^\circ$, $D=2,4,6$, and $L=2$. }
\label{fig:Energy}
\end{figure}
\begin{figure}
\includegraphics[width=0.45\textwidth]{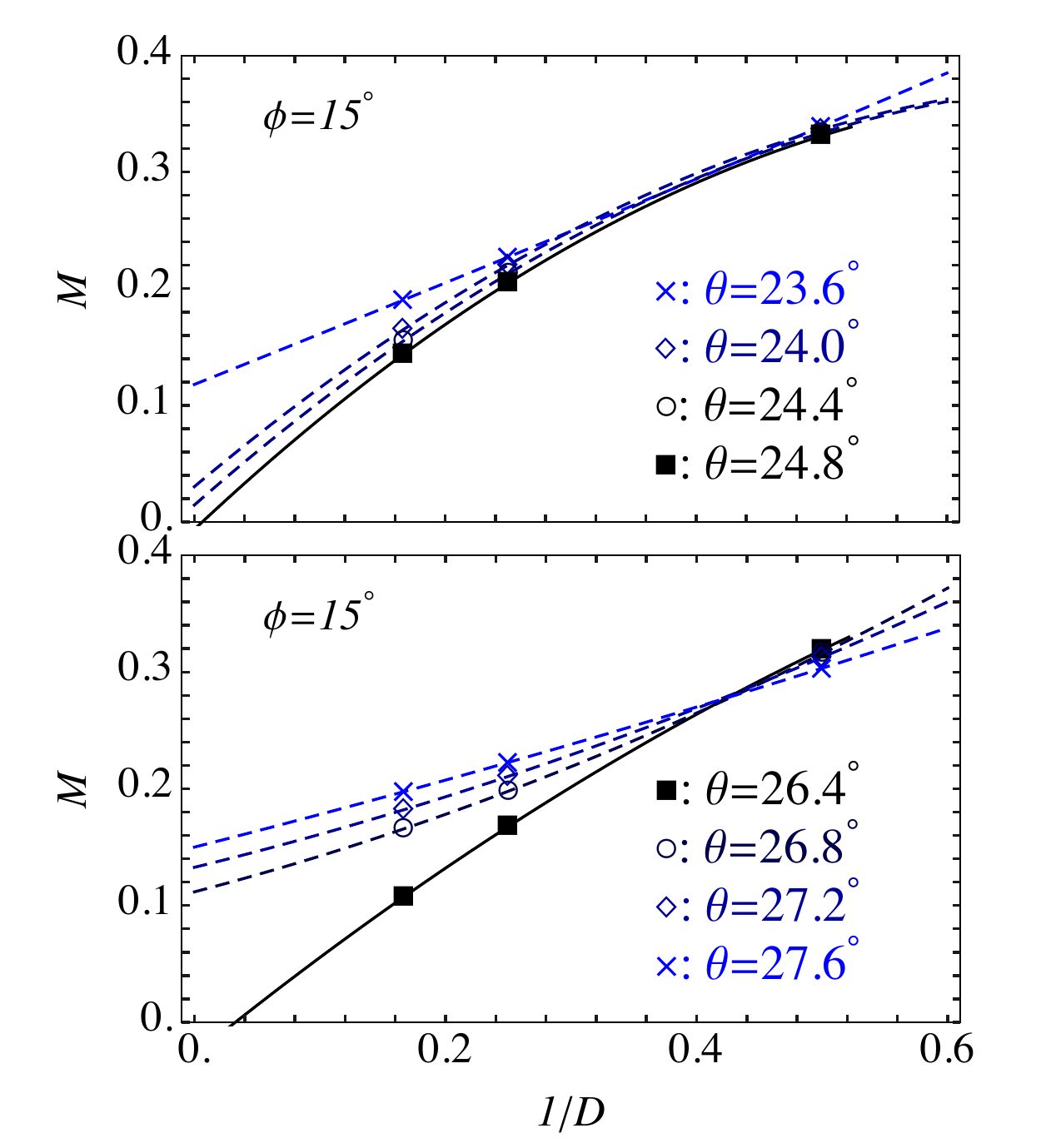}
\caption{ Extrapolations of $M$ in $1/D$ with $\phi=15^\circ$ and varied $\theta$. For $\theta\in [24.8^\circ, 26.4^\circ]$, $M$ is suppressed.}
\label{fig:trg2}
\end{figure}

 
The simple update and coarse-graining TRG steps are repeated until the average energy $E$ (Fig. \ref{fig:Energy}) is converged for given $D$. To obtain the phase boundary, we apply the finite-size extrapolations of $M$ using second-order polynomial fit in $1/D$ to infinite $D$~\cite{J1J2Jiang2,tnj1j22}. One example at $\phi=15^\circ$ is shown in Fig.~\ref{fig:trg2}. Suppression of the magnetization to zero as $D\rightarrow \infty$ suggests a quantum paramagnetic region. 

\subsection{Results for Finite Anisotropy}

\begin{figure}
\includegraphics[width=0.4\textwidth]{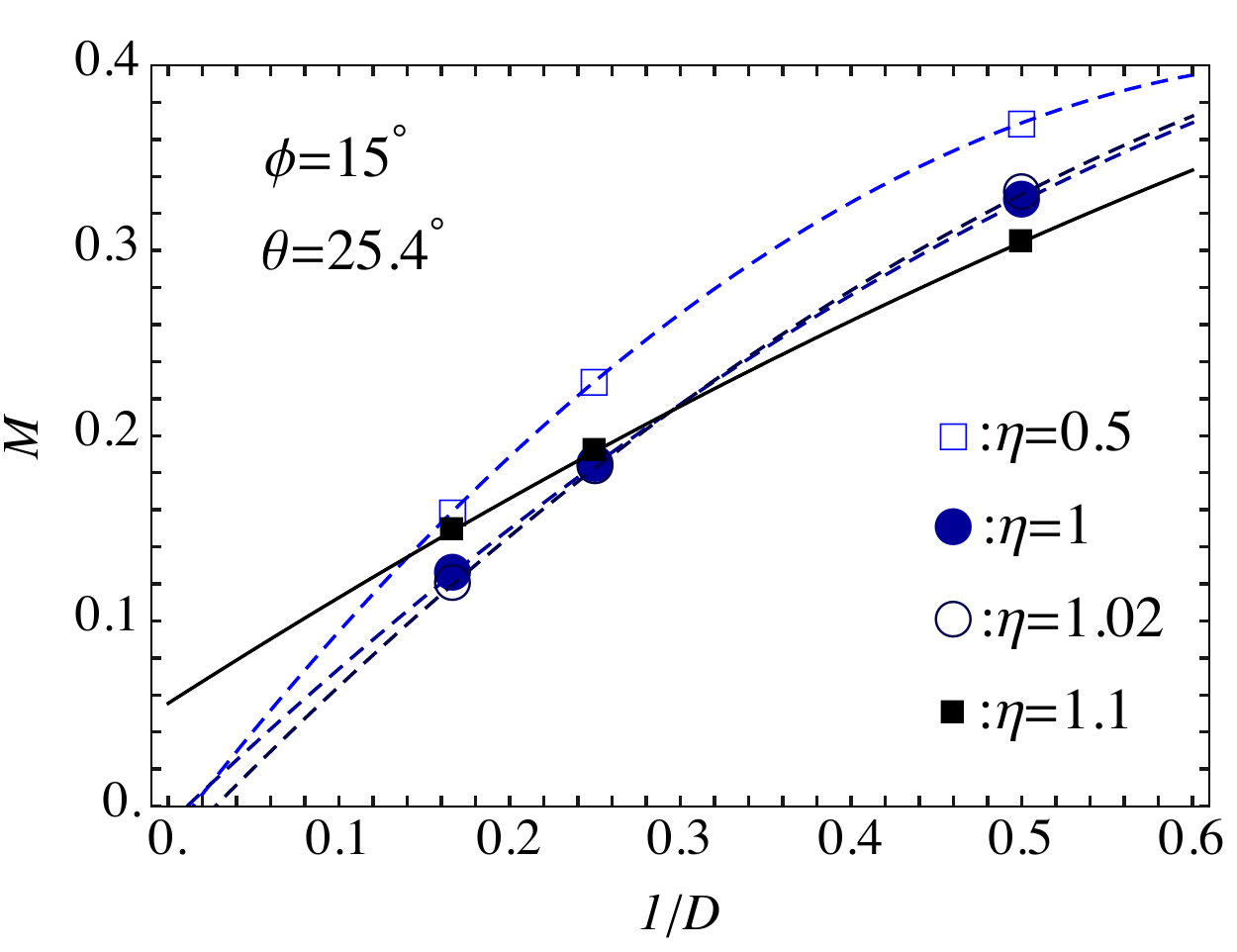}
\caption{ Extrapolation of $M$ in $1/D$ with $\phi=15^\circ$, $\theta=25.4^\circ$ and $\eta=0.5,1,1.02$, and $1.1$. }
\label{fig:eta}
\end{figure}

We apply the same extrapolations of $M$ for different anisotropy $\eta$ (Fig.~\ref{fig:eta}), which shows that the quantum paramagnetic region persists away from the Heisenberg limit $\eta=1$. Specifically, for $\eta< 1$, the quantum paramagnetic region remains robust for a large region, e.g., down to $\eta=0.5$. While for $\eta>1$, long range order is preferred when $\eta$ is increased to $\eta = 1.1$. This seems to suggest that the Heisenberg limit is close to the upper limit of the quantum paramagnetic region.

\end{document}